\documentclass[journal]{IEEEtran}
\usepackage{amsmath,amssymb,amsfonts}
\usepackage{tabularx}
\usepackage[utf8]{inputenc} 
\usepackage[T1]{fontenc}    
\usepackage{url}            
\usepackage{booktabs}       
\usepackage{amsfonts}       
\usepackage{nicefrac}       
\usepackage{microtype}      
\usepackage{graphicx}
\usepackage{float}
\restylefloat{table}
\usepackage{hyperref}
\usepackage{multicol}
\usepackage{caption}
\usepackage{subcaption}
\usepackage{amsmath}
\usepackage{algorithm}
\usepackage{algpseudocode}
\usepackage{tikz}
\usetikzlibrary{trees}
\usepackage{listings}
\usepackage{array}
\usepackage{colortbl,hhline}
\usepackage{color}
\usepackage{multirow}
\usepackage{xcolor}

\usepackage{textcomp}

\usepackage{xcolor,soul,framed} 

\usepackage[noadjust]{cite}

\usepackage[font=scriptsize]{caption}
\begin{document}
\bstctlcite{IEEEexample:BSTcontrol}
    \title{test}
\title{Coaching with PID Controllers: A Novel Approach for Merging Control with Reinforcement Learning}

\author{ Liping Bai \thanks{Nanjing Unversity of Posts and Telecommunications, College of Automation \& College of Artificial Intelligence, Nanjing, Jiangsu,210000 China email:zqpang@njupt.edu.cn}}
\maketitle
\begin{abstract}
We propose a Proportional Integral Derivative (PID) controller-based coaching scheme to expedite reinforcement learning (RL). Previous attempts to fuse classical control and RL are variations on imitation learning or human-in-the-loop learning. Also, in their approaches, the training acceleration comes with an implicit cap on what is attainable by the RL agent, therefore it is vital to have high-quality controllers. We ask if it is possible to accelerate RL with even a primitive hand-tuned PID controller, and we draw inspiration from the relationship between athletes and their coaches. At the top level of the athletic world, a coach's job is not to function as a template to be imitated after, but rather to provide conditions for the athletes to collect critical experiences. We seek to construct a coaching relationship between the PID controller and the RL agent, where the controller helps the agent experience the most pertinent states for the task. We conduct experiments in Mujoco locomotion simulation, but the setup can be easily converted into real-world circumstances. We conclude from the data that when the coaching structure between the PID controller and its respective RL agent is set at its goldilocks spot, the agent's training can be accelerated by up to 37\%, yielding uncompromised training results in the meantime. This is an important proof of concept that controller-based coaching can be a novel and effective paradigm for merging classical control with learning and warrants further investigations in this direction. All the code and data can be found at \href{https://github.com/BaiLiping/Coaching}{github/BaiLiping/Coaching}
\end{abstract}
\IEEEpeerreviewmaketitle
\section{Introduction}
\IEEEPARstart{L}{earning} for Dynamic Control is an emerging field of research located at the interaction between classic control and reinforcement learning (RL). Although RL community routinely generate jaw-dropping results that seem out of reach to the control community\cite{Andrychowicz2020LearningDI}\cite{Kalashnikov2018QTOptSD}\cite{Lee2020LearningQL}, the theories that undergird RL are as bleak as it was first introduced\cite{Bertsekas1996NeuroDynamicP}. Today, those deficiencies can be easily papered over by the advent of Deep Neural Networks (DNN) and ever faster computational capacities. For RL to reach its full potential, existing control theories and strategies have to be part of that new combined formulation.

There are three ways that classic control finds its way into RL. First, theorists who are well versed in optimization techniques and mathematical formalism can provide systematic perspectives to RL and invent the much needed analytical tools\cite{Han2020ActorCriticRL}\cite{Weinan2017APO}\cite{Dupont2019AugmentedNO}\cite{Betancourt2018OnSO}\cite{Nachum2020ReinforcementLV}. Second, system identification researchers are exploring all possible configurations to combine existing system models with DNN and its variations\cite{Hewing2020LearningBasedMP}\cite{Mohan2020EmbeddingHP}\cite{Lusch2018DeepLF}\cite{Bai2019DeepEM}\cite{BelbutePeres2020CombiningDP}. Third, proven controllers can provide data on successful control trajectories to be used in imitation learning, reverse reinforcement learning, and "human"-in-the-loop learning\cite{Knox2009InteractivelySA}\cite{Knox2010CombiningMF}\cite{Peng2018DeepMimicED}\cite{Peng2020LearningAR}\cite{Paine2018OneShotHI}.

Our approach is an extension of the third way of combining classing control with RL. Previous researches\cite{Xie2018LearningWT}\cite{Carlucho2017IncrementalQS}\cite{Pavse2020RIDMRI} are about making a functioning controller works better. To begin with, they require high-quality controllers, and the improvements brought about by the RL agents are merely icing on the cake. In addition, the controllers inadvertently impose limits on what can be achieved by the RL agents. If, unfortunately, a bad controller is chosen, then the RL training process would be hindered rather than expedited. We ask the question, can we speed up RL training with hand-tuned PID controllers, which are primitive but still captures some of our understanding of the system? This inquiry leads us to the relationship between athletes and their coaches. 

Professional athletes don't get where they are via trial-and-error. Their skillsets are forged through painstakingly designed coaching techniques. At the top level, a coach's objective is not to be a template for the athletes to imitate after, but rather is to facilitate data collection on critical states. Top athletes are not necessarily good coaches and vice versa. 

In our approach, the 'coaches' are PID controllers which we deliberately tuned to be barely functioning, as shown by Table \ref{score_compare}. Yet, even with such bad controllers, when appropriately structured, training acceleration is still observed in our experiments, as shown by Table \ref{episode_compare}. The critical idea of coaching is for the PID controllers to take over when the RL agents deviated from the essential states. Our approach differs from previous researches in one significant way: controllers' interventions and the rewards associated with such interventions are hidden from the RL agents. They are not part of the training data. We also restrain from reward engineering and leave everything as it is, other than the coaching intervention. This way, we can be confident that the observed acceleration does not stem from other alterations. The implementations would be detailed in subsequent sections.

\definecolor{airforceblue}{rgb}{0.36, 0.54, 0.66}
\definecolor{beaublue}{rgb}{0.74, 0.83, 0.9}

\begin{table}[H]
\footnotesize
\caption{Performance Comparison between PID controller and its respective RL agent. We interfaced with Mujoco simulation through OpenAI GYM, and every simulated environment comes with predetermined maximum episode steps. The scores achieved by the RL agents would probably be high if not for this reason.}
\label{score_compare}
\centering
\begin{tabular}{ cccc }
\rowcolor{airforceblue}
Environment &   PID Controller &RL Agent &PID\slash RL \\
Inverted Pendulum &  240 & 1000&  24.0\%\\
\rowcolor{beaublue}

Double Pendulum &  1107 & 9319& 11.9\%\\

Hopper &  581 & 989& 58.7\%\\
\rowcolor{beaublue}
Walker &  528 & 1005& 52.5\%\\
\end{tabular}
\end{table}

\begin{table}[H]
\scriptsize
\caption{Comparison Between Agents Trained With and Without PID Controller Coaching. Even though the PID controllers are less capable than the eventual RL agent, they are still useful and can accelerate the RL agent training. There two measures we used to gauge training acceleration. The first is five consecutive wins, and the second is the scoring average. The "win" is a predetermined benchmark. }
\label{episode_compare}
\centering
\begin{tabular}{ cccccc }
\rowcolor{airforceblue}

Environment & Target & Measure  &  With PID  & Without  & Percentage\\
\rowcolor{airforceblue}

   Name     & Score  &              & Coaching  & Coaching  & Increase \\
\hline
Inverted & 800& Win Streak & 100 & 160&  37.5\% \\
Pendulum & &Average  & 104 &  159&  34.6\%\\
\rowcolor{beaublue}
Double & 5500& 5 Wins & 908 & 1335&  31.9\%\\
\rowcolor{beaublue}
Pendulum & &Average & 935 &  1370&  29.9\%\\
Hopper & 800& 5 Wins & 2073 & 2851&  27.3\%\\
       & &Average  & 2155 &  2911&  25.9\%\\
\rowcolor{beaublue}
Walker & 800& 5 Wins & 4784 & 5170&  7.5\%\\
\rowcolor{beaublue}
       & &Average  & 5659 &  7135&  20.7\%\\

\end{tabular}
\end{table}

In section II, we present the idea of controller-based coaching. In section III, we present the results of our experiments and their detailed implementations. We conclude what we have learned and layout directions for further research in section IV.

\section{Controller Based Coaching}
\begin{figure}[H]
    \centering
    \includegraphics[width=0.5\textwidth]{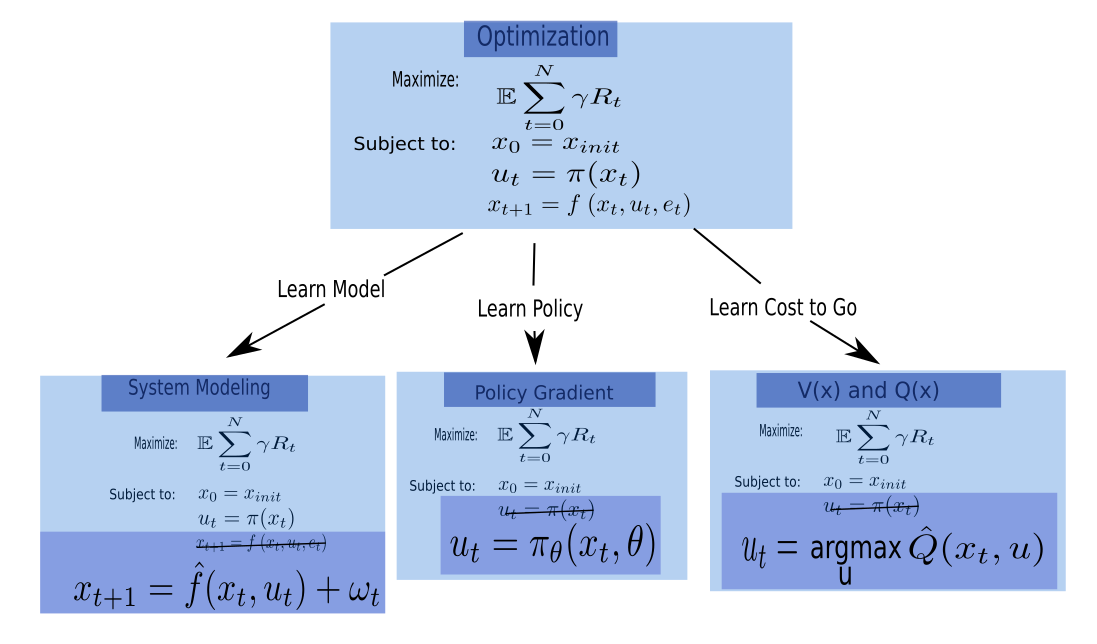}
    \caption{From Optimization to Learning. Model-Based or Model-Free learning refers to whether or not learning is used to approximate the system dynamics function. If there is an explicit action policy, it is called on-policy learning. Otherwise, the optimal action would be implicitly captured by the Q value function, and that would be called off-policy learning instead. Importance sampling allows "limited off-policy learning" capacity, which enables data reuse in a trusted region. Online learning means interleaving data collection and iterative network parameters update. Offline learning means the data is collected in bulk first, and then the network parameters are set with regression computation. Batch learning, as the name suggested, is in between online and offline learning. An agent would first generate data that fill its batch memory and then sample from the batch memories for iterative parameter update. New data would be generated with the updated parameters to replace older data in the memory. This taxonomy is somewhat outdated now. When Richard Sutton wrote his book, the algorisms he had in mind fall nicely into various categories. Today, however, the popular algorisms would combine more than one route to derive superior performance and can't be pigeonholed.}
    \label{fig:rl}
\end{figure}

Reinforcement Learning is the process of cycling between interaction with an environment and refinement of the understanding of that environment. RL agents methodically extract information from experiences, gradually bounding system models, policy distributions, or cost-to-go approximations to maximize the expected rewards along a trajectory, as shown by Figure\ref{fig:rl} which is an adaption of Benjamin Recht's presentation\cite{Recht2018ATO}. 

\begin{figure}
\centering
\begin{subfigure}{0.2\textwidth}
  \centering
  \includegraphics[width=\linewidth]{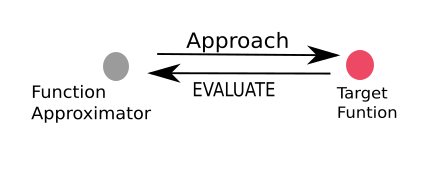}
  \caption{With Known Evaluation Function}
  \label{fig:known}
\end{subfigure}%
\begin{subfigure}{.3\textwidth}
  \centering
  \includegraphics[width=\linewidth]{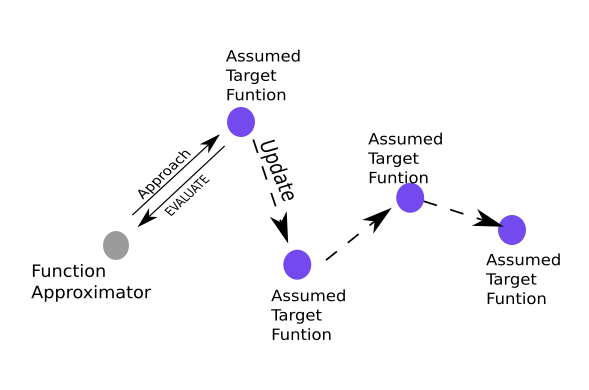}
  \caption{Bootstrap}
  \label{fig:unknown}
\end{subfigure}
\label{fig:bootstrap}
\end{figure}

A fundamental concept for RL is convergence through bootstrap. Instead of asymptotically approaching a known target function\ref{fig:known}, bootstrap methods approach an assumed target first and then update the target assumption based on collected data\ref{fig:unknown}. When evaluating estimation functions with belief rather than of the real value, things could just run around in circles and never converge. Without any guidance, the RL agent would have just explored all the possible states, potentially resulting in this unstable behavior. 

One method to produce more efficient exploration and avoid instability is to give more weight to critical states. Not all observational states are created equal. Some are vital, while others have nothing to do with the eventual objective. For instance, in the inverted pendulum task, any states outside of the Lyapunov stability bounds should be ignored since they can't be properly controlled anyway. 

There are statistical techniques to distinguish critical states from the non-essential ones, and imitation learning works by marking crucial states with demonstrations. However, the former approach is hard to implement, and the latter one requires high-quality controllers. Our proposed controller-based coaching method is easy to implement and does not have stringent requirements on the controllers it uses.

Controller-based coaching works by adjusting the trajectory of the RL agent and avoid wasting valuable data collection cycle on the states that merely lead to eventual termination. When the agent is about to deviate from essential states, the controller will apply a force to nudge the agent back to where it should be, much like a human coach course-corrects on behalf of the athlete. Crucially, the agent is oblivious to this intervention step, and it would not be part of the agent's data. Even if the controller didn't adjust the agent to where it should be, it would not have mattered since it is unaware of it because it is a high-quality controller. On the other hand, if the controller successfully adjusts the trajectory, the RL agent's next data collection cycle will be spent in a critical state. We test our approach on four mujoco locomotion environments as a proof of concept, and in all four experiments, the hypothesized acceleration on RL training is observed.

\section{Experiment Setup}
Mujoco physics engine\cite{6386109}, is one of many such simulation tools. We interface with it through a python wrapper provided by the OpenAI Gym\cite{Brockman2016OpenAIG} team. We choose four environments for our experiments: inverted pendulum, double inverted pendulum, hopper, and walker. Every environment comes with a set of predetermined rewards and maximum episode steps. We did not tinker with those parameters. The only change we made to each environment is a controller-based coach ready to intervene when the agent steps out of the predetermined critical states.

We use tensorforce's\cite{tensorforce} implementation of RL agents, specifically the Proximal Policy Optimization (PPO) agent because the learning curves generated by PPO agent are smoother, as shown by the spinning up\cite{SpinningUp2018} team. Our paper aims to indicate the controller-based coaching method's feasibility, and a smoother curve makes things easier. 

In this paper, human judgment is the basis for the determination of critical states. In future works, we would like to provide a more systematic process for critical states' demarkation. We will provide our reasonings when we discuss our experiments in each environment.

Our experiments' code and data can be accessed via \href{https://github.com/BaiLiping/Coaching}{this deposite}. The folders are named after each environment, and in each folder, you will find a data record, an RL agent record, a agent.json file that indicates all the hyperparameters of the RL agent, a Normal file that trains an RL agent with the normal method, a Coached file which trains an agent based on the controller based coaching method, a PID file which is the PID coach. In the PIDvsRL folder, you will find files that generate all the plots shown in the following section.
\subsection{Inveretd Pendulum}
The observation space of the inverted pendulum environment is: [Cart Position on X-axis, Cart Velocity, Pole Angle, Pole Angular Velocity]. The continuous action space is an action ranging from -10 to 10, where -10 means the actuator moves the cart to the left with maximum power and 10 means the actuator moves the cart to the right with full force. The maximum episode step number is 1000. The terminal state for the inverted pendulum is an angle of absolute value greater than 0.2 radius. The reward is 1 for each non-terminal step and 0 for the terminal step. 

\begin{figure}
\centering
\begin{subfigure}{0.24\textwidth}
  \centering
  \includegraphics[width=\linewidth]{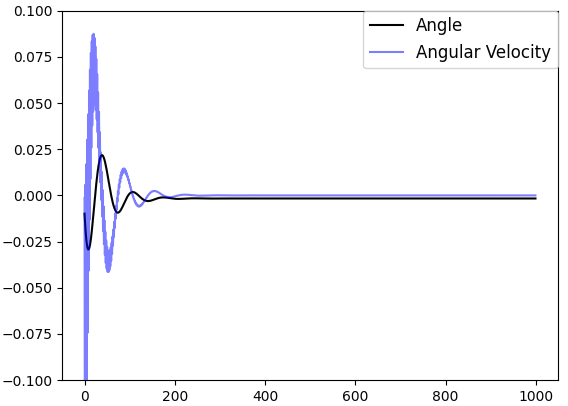}
\end{subfigure}
\hfill
\begin{subfigure}{.24\textwidth}
  \centering
  \includegraphics[width=\linewidth]{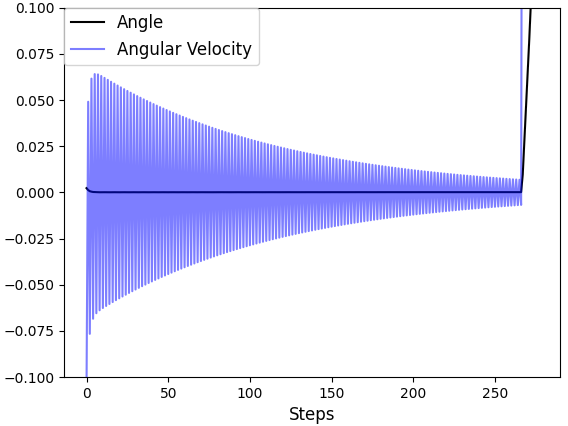}
\end{subfigure}
\caption{Inverted Pendulum system controlled by RL agent and its PID coach. The left plot is RL agent and the right plot is the PID controller. The maximum episode steps are 1000 and each step without termination is scored 1 point. The average score achieved by the RL agent is 1000 and the average score achieved by the PID controller is 240.}
\label{fig:ip}
\end{figure}

Figure\ref{fig:ip} shows how the RL agent and the PID controller manage the pole angle and the angular velocity. The left plot is the RL agent, and the right plot is the PID controller coach. While the PID controller tries hard to converge to zero, eventually, there would be too much accumulation on the integral term, and the equilibrium breaks down. The average score achieved by the RL agent is 1000, and the average score achieved by the PID controller is 240.

Based on our observation of the system, we decided to put the boundary between critical and noncritical states on the angular velocity of the absolute value of 0.4. The RL agent is free to explore with the angular velocity with an absolute value smaller than 0.4, but once it goes above this bound, the PID controller will kick in, trying to decrease the velocity back to the 0.4 bound.

\begin{figure}
     \centering
      \includegraphics[width=0.5\textwidth]{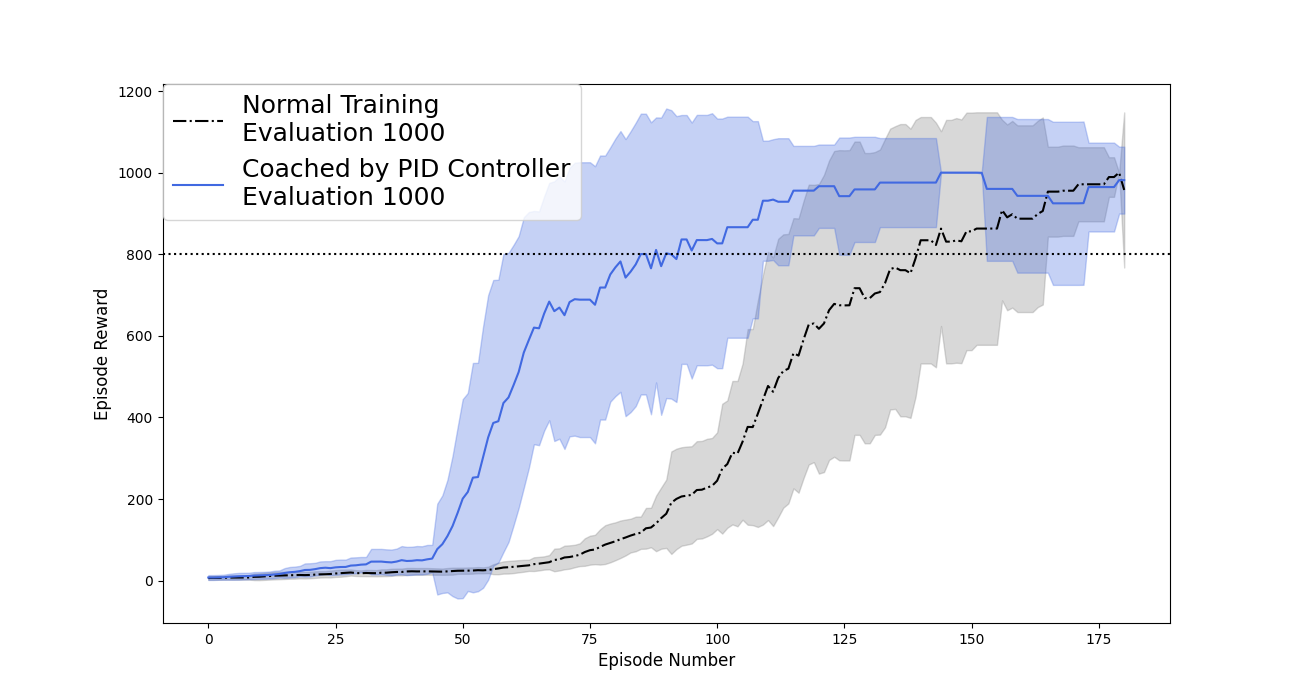}
      \caption{Inverted Pendulum Coaching Result}
      \label{fig:ip_result}

\end{figure}

The experiment result is presented by Figure\ref{fig:ip_result}. The black line indicates the agent trained normally, and the blue line indicates the agent trained with a PID coach. A win is set to be with a score greater than 800. It takes the normal method 160 episodes to get five consecutive wins, and it takes the controller-based coaching method 100 episodes to do the same. As measured by averaging over 10 episodes, it takes the normal method 159 episodes to go beyond 800, and it takes controller-based coaching method 104 episodes to do the same. The acceleration is roughly 35\%. The agents trained with both methods pass the evaluation, and their respective average scores are presented in the upper left corner.

\subsection{Inverted Double Pendulum}
The inverted double pendulum has observation space of the following: [x position of the cart, sin($\theta_1$), sin($\theta_2$),cos($\theta_1$),cos($\theta_2$),velocity of x, angular velocity of $\theta_1$, angular velocity of $\theta_2$, constraint force on x, constraint force on $\theta_1$, constraint force on $\theta_2$]. $\theta_1$ and $\theta_2$ are the angles of the upper and lower pole perspectively. The action space for Inverted Double Pendulum is an action ranging from -10 to 10, where -10 means the actuator moves the cart to the left with maximum power and 10 means the actuator moves the cart to the right with maximum power. A score of roughly 10 points is assigned to non-terminal states, based on the velocity on the x-axis. The detailed formular for score computation can be found at the OpenAI site.

\begin{figure}
\centering
\begin{subfigure}{0.24\textwidth}
  \centering
  \includegraphics[width=\linewidth]{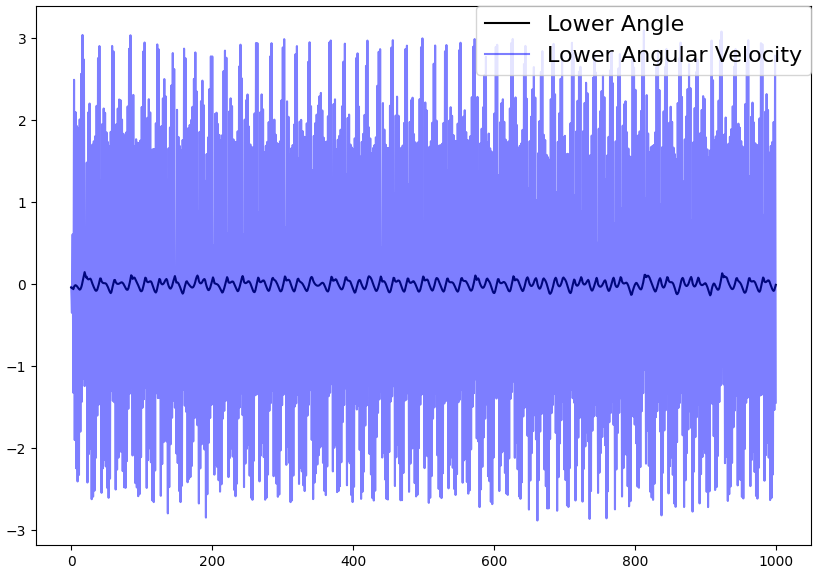}
\end{subfigure}%
\hfill
\begin{subfigure}{.24\textwidth}
  \centering
  \includegraphics[width=\linewidth]{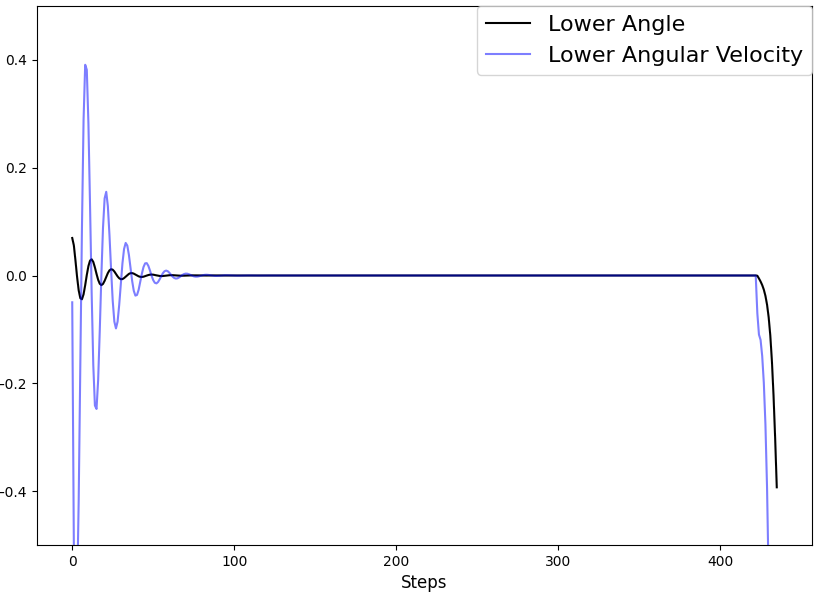}
\end{subfigure}
\caption{Inverted Double Pendulum system controlled by RL agent and its PID coach. The left plot is RL agent and the right plot is the PID controller. The maximum episode steps are 1000 and each step without termination is scored at around 10 points, based on velocity on the x axis. The average score achieved by the RL agent is 9319 and the average score achieved by the PID controller is 1107. }
\label{fig:double}
\end{figure}

Figure\ref{fig:double} shows how the RL agent and the PID controller manage the lower pole angle and its angular velocity. The left plot is the RL agent and the right plot is the PID controller coach. The RL agent seems to settle on continuously oscillating from the left limit to the right limit until the maximum episode steps are reached. The PID controller functions well until the equilibrium breaks down with too much accumulation on the integral term. The average score achieved by the RL agent is 9319, and the average score achieved by the PID controller is 1107.

Based on our observation of the system, we decided to put the boundary between critical and noncritical states on the lower angle of the absolute value of 0.2. The RL agent is free to explore with the lower angle with an absolute value smaller than 0.2, but once it goes above this bound, the PID controller will kick in, trying to nudge the lower angle back to the 0.2 bound.

\begin{figure}
     \centering
      \includegraphics[width=0.5\textwidth]{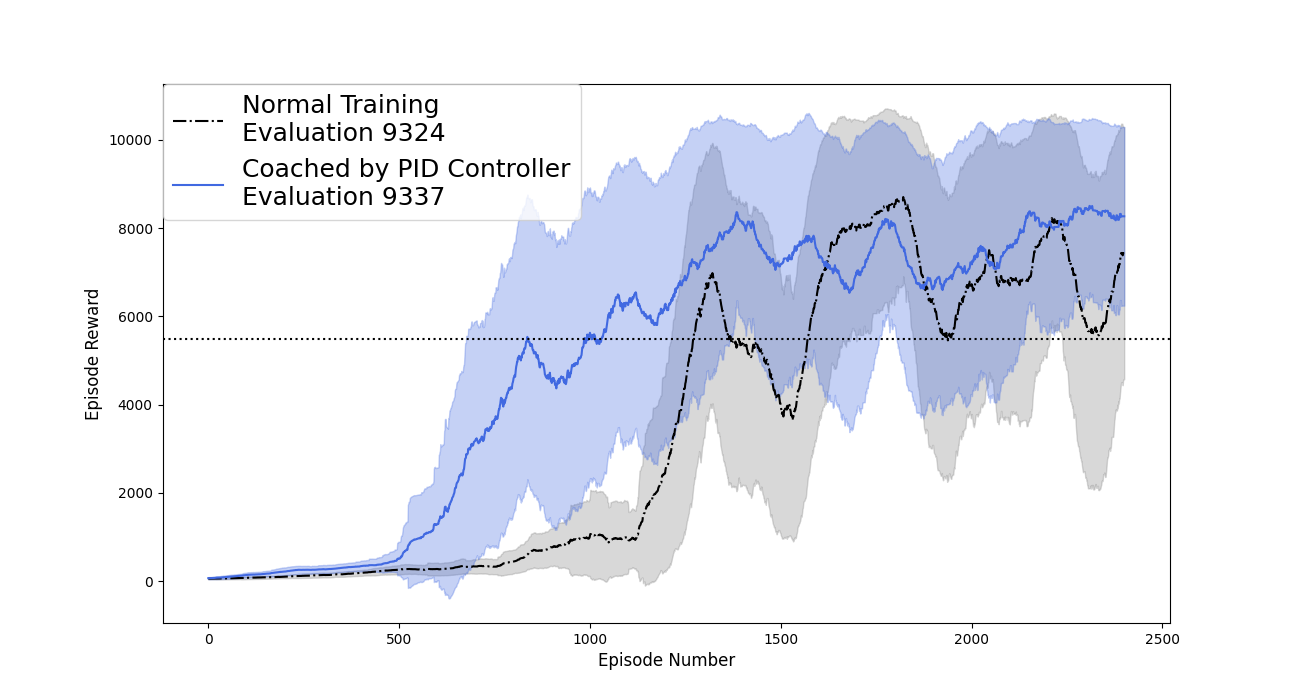}
      \caption{Double Inverted Pendulum Coaching Result.}
      \label{fig:double_result}
\end{figure}

The experiment result is presented by Figure\ref{fig:double_result}. The black line indicates the agent trained normally, and the blue line indicates the agent trained with a PID coach. A win is set to be with a score greater than 5500. It takes the normal method 1335 episodes to get five consecutive wins, and it takes controller-based coaching method 908 episodes to do the same. As measured by averaging over 100 episodes, it takes normal method 1370 episodes to go beyond 5500, and it takes controller-based coaching method 935 episodes to do the same. The acceleration is roughly 30\%. The agents trained with both methods pass the evaluation, and their respective average scores are presented in the upper left corner.

\subsection{Hopper}
The observation space of hopper is the following vector: [z position, y position, thigh joint angle, leg joint angle, foot joint angle, velocity at x-axis, velocity at z-axis, velocity at y-axis, angular velocity of thigh joint, angular velocity of leg joint, angular velocity of foot joint]. The hopper's action space is three continuous action choices for three actuators [thigh actuator, leg actuator, foot actuator]. The range of actuators is -10, which means applying force towards the negative direction with maximum power, and 10, which means applying force towards the positive direction with maximum power. The terminal states for hopper are when the absolute y position is greater than 0.2.

Based on our observation of the system, we decided to put the boundary between critical and noncritical states on the y axis velocity of the absolute value of 0.3. The RL agent is free to explore with the y axis velocity with an absolute value smaller than 0.3, but once it goes above this bound, the PID controller will kick in, trying to decrease the y axis velocity back to the 0.3 bound.

\begin{figure}
\centering
\begin{subfigure}{0.24\textwidth}
  \centering
  \includegraphics[width=\linewidth]{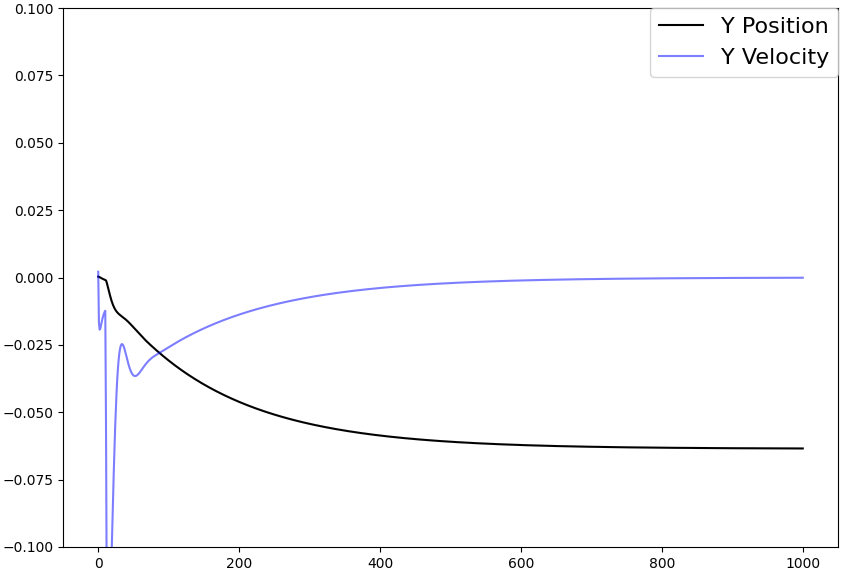}
\end{subfigure}
\hfill
\begin{subfigure}{.24\textwidth}
  \centering
  \includegraphics[width=\linewidth]{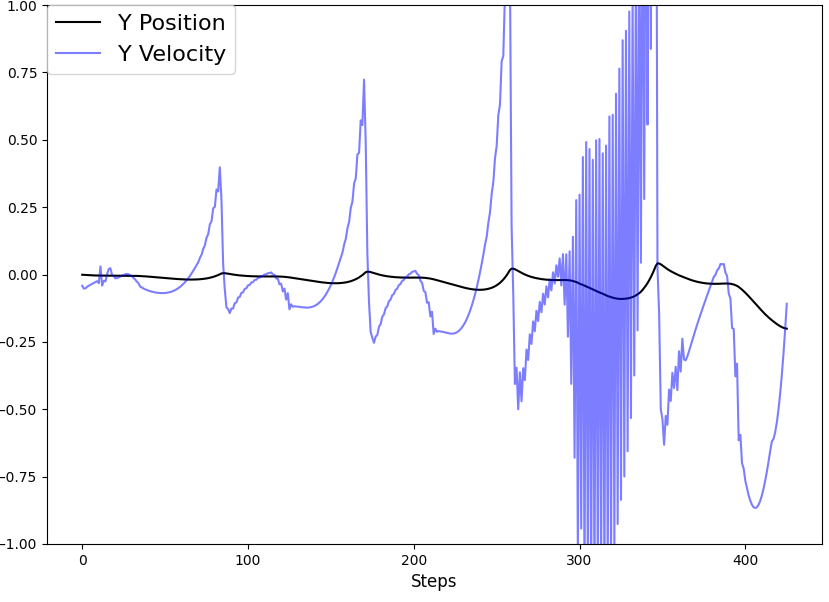}
\end{subfigure}
\caption{Hopper system controlled by RL agent and its PID coach. The left plot is RL agent and the right plot is the PID controller. The maximum episode steps are 1000 and each step without termination is scored at around 1 point, depending on the x-axis velocity. The average score achieved by the RL agent is 989 and the average score achieved by the PID controller is 581.}
\label{fig:hopper}
\end{figure}

Figure\ref{fig:hopper} shows how the RL agent and the PID controller manage they position and its velocity. The left plot is the RL agent and the right plot is the PID controller coach. The RL agent seems to settle on doing nothing until the maximum episode steps are reached. The PID controller can only manage they position into an oscillation with ever-increasing magnitude. The average score achieved by the RL agent is 989, and the average score achieved by the PID controller is 581.

\begin{figure}
     \centering
      \includegraphics[width=0.5\textwidth]{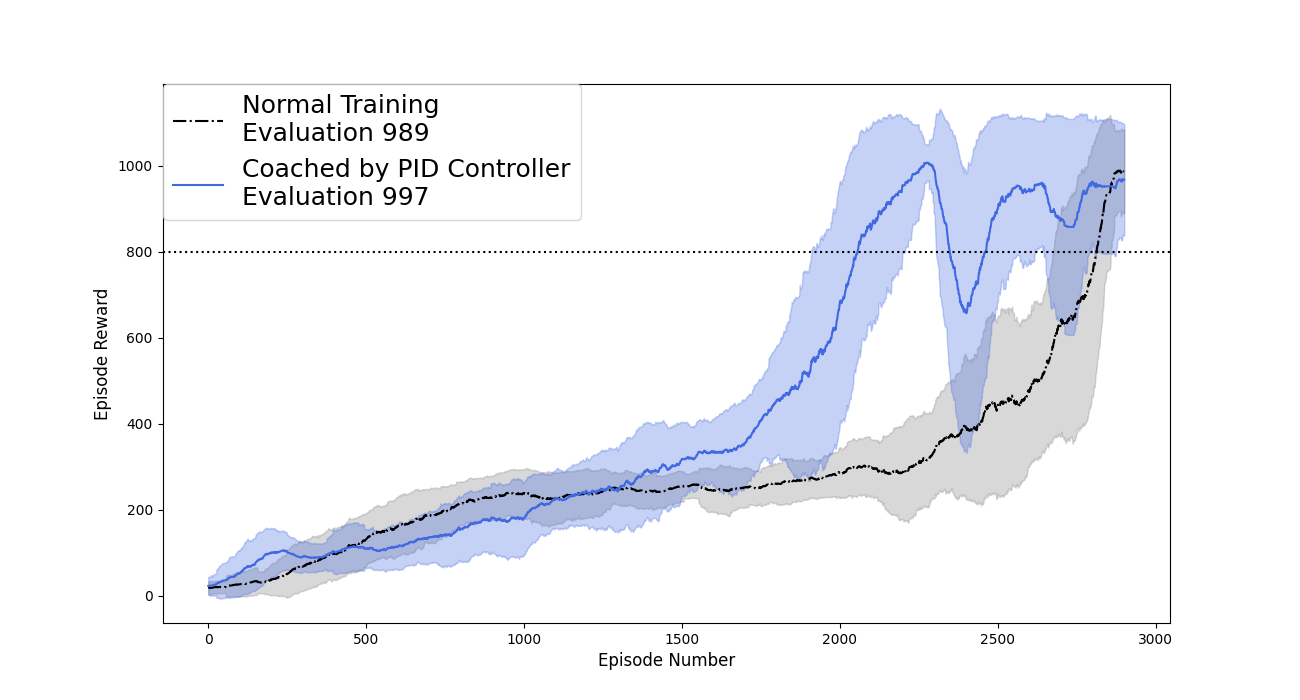}
      \caption{Hopper Coaching Result. Benchmarked against training without coaching, indicated by the black dotted line.}
      \label{fig:hopper_result}
\end{figure}

The experiment result is presented by Figure\ref{fig:hopper_result}. The agent trained normally is indicated by the black line, and the agent trained with PID coach is indicated by the blue line. A win is set to be with a score greater than 800. It takes the normal method 2851 episodes to get five consecutive wins, and it takes controller-based coaching method 2073 episode to do the same. As measured by averaging over 100 episodes, it takes the normal method 2911 episodes to go beyond 800, and it takes the controller-based coaching method 2155 episodes to do the same. The acceleration is roughly 25\%. The agents trained with both methods pass the evaluation, and their respective average scores are presented in the upper left corner.

\subsection{Walker}
The walker system is just a two-legged hopper. The observation space is the same as listed in the hopper environment but for both legs. The terminal state is when the z position falls below 0.8.

Based on our observation of the system, we decided to put the boundary between critical and noncritical states on the y axis velocity of the absolute value of 1. The RL agent is free to explore with the y axis velocity with an absolute value smaller than 1, but once it goes above this bound, the PID controller will kick in, trying to decrease the y axis velocity back to the 1 bound.

\begin{figure}
\centering
\begin{subfigure}{0.24\textwidth}
  \centering
  \includegraphics[width=\linewidth]{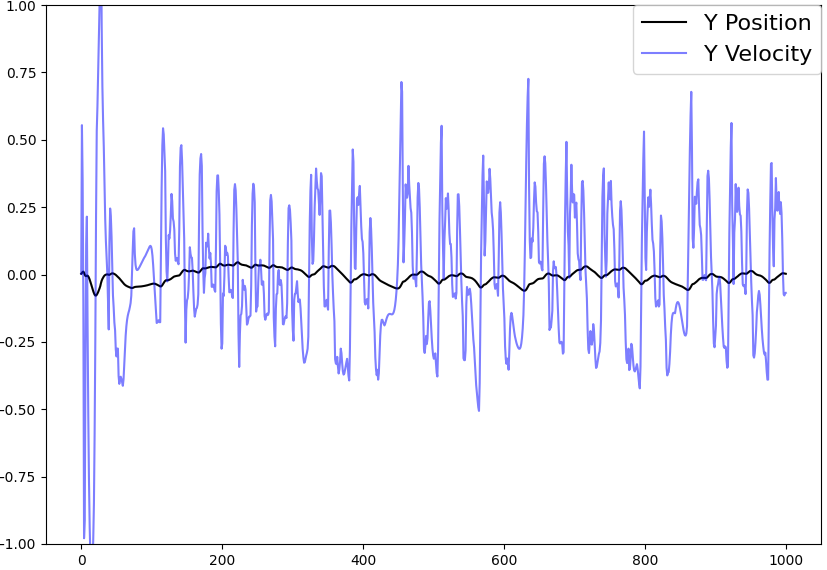}
\end{subfigure}
\hfill
\begin{subfigure}{.24\textwidth}
  \centering
  \includegraphics[width=\linewidth]{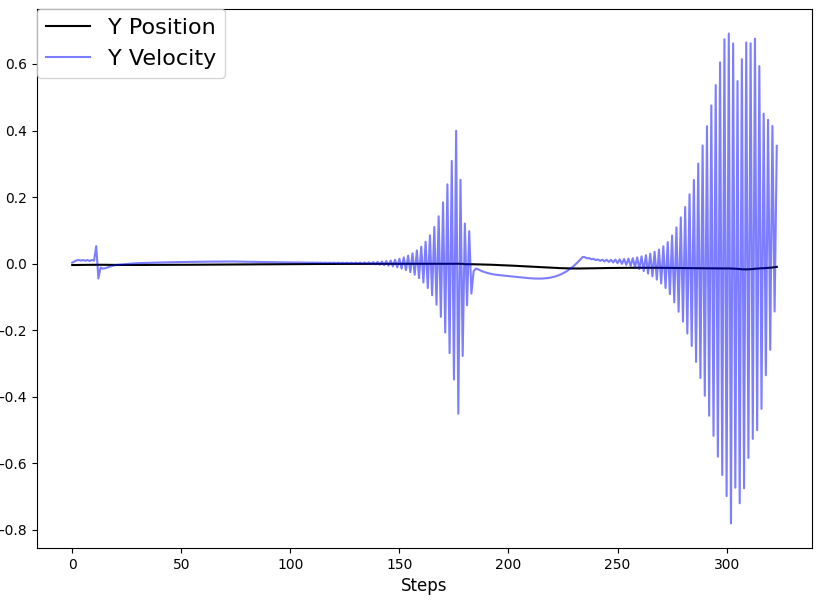}
\end{subfigure}
\caption{Walker system controlled by RL agent and its PID coach. The left plot is RL agent and the right plot is the PID coach. The maximum episode steps are 1000 and each step without termination is scored at around 1 point, depending on the x-axis velocity. The average score achieved by the RL agent is 1005 and the average score achieved by the PID controller is 528.}
\label{fig:walker}
\end{figure}

Figure\ref{fig:walker} shows how the RL agent and the PID controller manage the y position and its velocity. The left plot shows RL agent can handle the oscillation of y velocity in between the 0.5 bound, but the PID controller is incapable of it, as indicated by the right plot. The PID controller can only manage the y position into an oscillation with ever-increasing magnitude. The average score achieved by the RL agent is 1005, and the average score achieved by the PID controller is 528.

\begin{figure}
     \centering
      \includegraphics[width=0.5\textwidth]{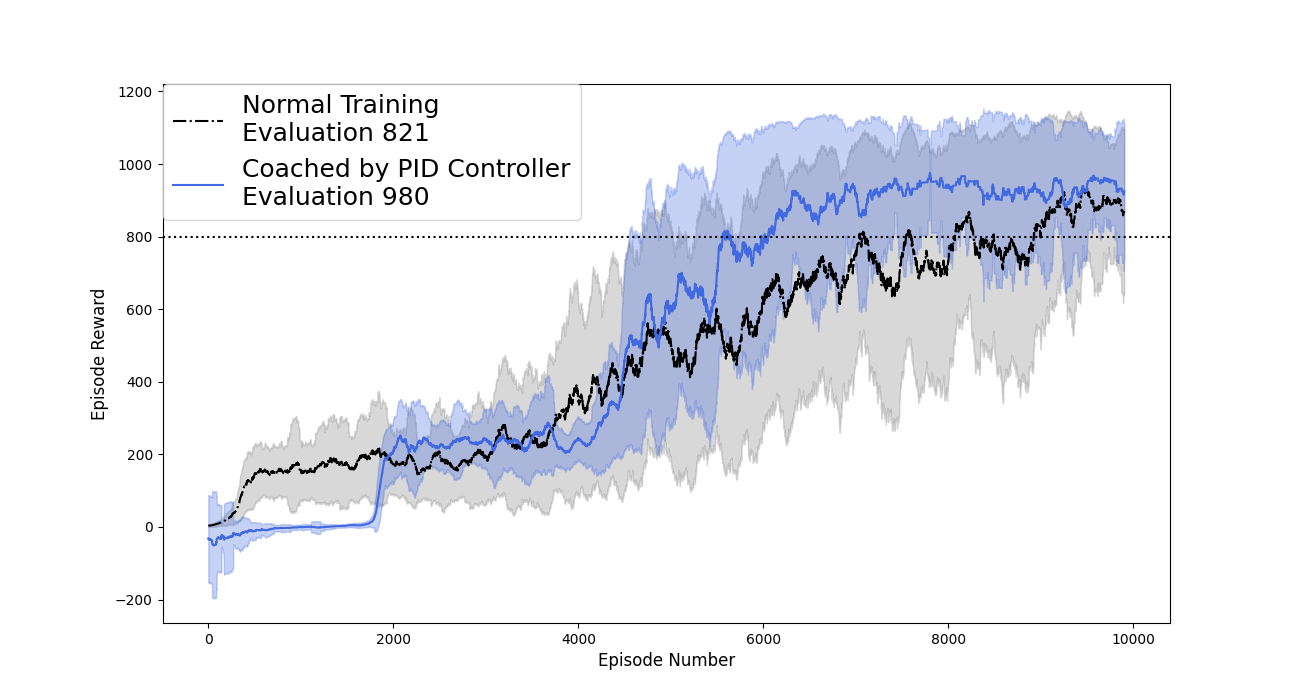}
      \caption{Hopper Coaching Result. Benchmarked against training without coaching, indicated by the black dotted line.}
      \label{fig:walker_result}
\end{figure}

The experiment result is presented by Figure\ref{fig:walker_result}. The black line indicates the agent trained normally, and the blue line indicates the agent trained with a PID coach. A win is set to be with a score greater than 800. It takes the normal method 5170 episodes to get five consecutive wins, and it takes controller-based coaching method 4784 episode to do the same. As measured by averaging over 100 episodes, it takes normal method 7135 episodes to go beyond 800, and it takes controller-based coaching method 5659 episodes to do the same. The acceleration is roughly 10\%. The agents trained with both methods pass the evaluation, and their respective average scores are presented in the upper left corner.

\section{Conclution}

In this paper, we present the controller based coaching approach for accelerated RL training. Unlike previous attempts for using the controller as a guide to RL agent, our method can work with the most primitive PID controllers. In all our experiments, the PID controllers are barely functioning, yet acceleration is still observed. We ascribe this to the fact that the coach intervention step is not part of the RL agent record. Therefore, even if the coach failed to do its job, it would not have worsened the RL training process. Yet when the coach does its job, accelerated data collection on the critical states is achieved. 

Next step, we plan to implement the controller-based coaching idea to the deep drone acrobat project \cite{Kaufmann2020DeepDA}. The drones are trained in simulation, and we think our approach can significantly accelerate their training. In future works, we want to provide a theoretical basis for distinguishing between critical and noncritical states, as opposed to base solely on human judgment, as we did here. 

We believe our research opens the door to a rich reservoir of potential research on coaching RL agents with controllers. Human achieves superhuman feats not because of talent, but because of the meticulously engineered coaching tactics. Current research in RL focuses solely on the "athlete" side of the equation, i.e., building an efficient RL agent, but we feel "coach" is as important, if not more so.  For instance, a tennis coach will challenge athletes, pushing them to experience situations that are hard to encounter. A controller can function as a coach and pushes the RL agent into states that are hard to access. Controller-based coaching can be an effective way to merge controllers with RL agents. 

\bibliographystyle{IEEEtran}
\bibliography{Bibliography}

\begin{thebibliography}{10}
\providecommand{\url}[1]{#1}
\csname url@rmstyle\endcsname
\providecommand{\newblock}{\relax}
\providecommand{\bibinfo}[2]{#2}
\providecommand\BIBentrySTDinterwordspacing{\spaceskip=0pt\relax}
\providecommand\BIBentryALTinterwordstretchfactor{4}
\providecommand\BIBentryALTinterwordspacing{\spaceskip=\fontdimen2\font plus
\BIBentryALTinterwordstretchfactor\fontdimen3\font minus
  \fontdimen4\font\relax}
\providecommand\BIBforeignlanguage[2]{{%
\expandafter\ifx\csname l@#1\endcsname\relax
\typeout{** WARNING: IEEEtran.bst: No hyphenation pattern has been}%
\typeout{** loaded for the language `#1'. Using the pattern for}%
\typeout{** the default language instead.}%
\else
\language=\csname l@#1\endcsname
\fi
#2}}

\bibitem{Andrychowicz2020LearningDI}
O.~M. Andrychowicz, B.~Baker, M.~Chociej, R.~J{\'o}zefowicz, B.~McGrew, J.~W.
  Pachocki, A.~Petron, M.~Plappert, G.~Powell, A.~Ray, J.~Schneider, S.~Sidor,
  J.~Tobin, P.~Welinder, L.~Weng, and W.~Zaremba, ``Learning dexterous in-hand
  manipulation,'' \emph{The International Journal of Robotics Research},
  vol.~39, pp. 20 -- 3, 2020.

\bibitem{Kalashnikov2018QTOptSD}
D.~Kalashnikov, A.~Irpan, P.~Pastor, J.~Ibarz, A.~Herzog, E.~Jang, D.~Quillen,
  E.~Holly, M.~Kalakrishnan, V.~Vanhoucke, and S.~Levine, ``Qt-opt: Scalable
  deep reinforcement learning for vision-based robotic manipulation,''
  \emph{ArXiv}, vol. abs/1806.10293, 2018.

\bibitem{Lee2020LearningQL}
J.~Lee, J.~Hwangbo, L.~Wellhausen, V.~Koltun, and M.~Hutter, ``Learning
  quadrupedal locomotion over challenging terrain,'' \emph{Science Robotics},
  vol.~5, 2020.

\bibitem{Bertsekas1996NeuroDynamicP}
D.~Bertsekas and J.~Tsitsiklis, ``Neuro-dynamic programming,'' in
  \emph{Encyclopedia of Machine Learning}, 1996.

\bibitem{Han2020ActorCriticRL}
M.~Han, L.~Zhang, J.~Wang, and W.~Pan, ``Actor-critic reinforcement learning
  for control with stability guarantee,'' \emph{IEEE Robotics and Automation
  Letters}, vol.~5, pp. 6217--6224, 2020.

\bibitem{Weinan2017APO}
E.~Weinan, ``A proposal on machine learning via dynamical systems,'' 2017.

\bibitem{Dupont2019AugmentedNO}
E.~Dupont, A.~Doucet, and Y.~Teh, ``Augmented neural odes,'' in \emph{NeurIPS},
  2019.

\bibitem{Betancourt2018OnSO}
M.~Betancourt, M.~I. Jordan, and A.~Wilson, ``On symplectic optimization,''
  \emph{arXiv: Computation}, 2018.

\bibitem{Nachum2020ReinforcementLV}
O.~Nachum and B.~Dai, ``Reinforcement learning via fenchel-rockafellar
  duality,'' \emph{ArXiv}, vol. abs/2001.01866, 2020.

\bibitem{Hewing2020LearningBasedMP}
L.~Hewing, K.~P. Wabersich, M.~Menner, and M.~N. Zeilinger, ``Learning-based
  model predictive control: Toward safe learning in control,'' 2020.

\bibitem{Mohan2020EmbeddingHP}
A.~Mohan, N.~Lubbers, D.~Livescu, and M.~Chertkov, ``Embedding hard physical
  constraints in neural network coarse-graining of 3d turbulence.''
  \emph{arXiv: Computational Physics}, 2020.

\bibitem{Lusch2018DeepLF}
B.~Lusch, J.~N. Kutz, and S.~Brunton, ``Deep learning for universal linear
  embeddings of nonlinear dynamics,'' \emph{Nature Communications}, vol.~9,
  2018.

\bibitem{Bai2019DeepEM}
S.~Bai, J.~Z. Kolter, and V.~Koltun, ``Deep equilibrium models,'' \emph{ArXiv},
  vol. abs/1909.01377, 2019.

\bibitem{BelbutePeres2020CombiningDP}
F.~de~Avila Belbute-Peres, T.~D. Economon, and J.~Z. Kolter, ``Combining
  differentiable pde solvers and graph neural networks for fluid flow
  prediction,'' \emph{ArXiv}, vol. abs/2007.04439, 2020.

\bibitem{Knox2009InteractivelySA}
W.~B. Knox and P.~Stone, ``Interactively shaping agents via human
  reinforcement: the tamer framework,'' in \emph{K-CAP '09}, 2009.

\bibitem{Knox2010CombiningMF}
------, ``Combining manual feedback with subsequent mdp reward signals for
  reinforcement learning,'' in \emph{AAMAS}, 2010.

\bibitem{Peng2018DeepMimicED}
X.~Peng, P.~Abbeel, S.~Levine, and M.~V.~D. Panne, ``Deepmimic: Example-guided
  deep reinforcement learning of physics-based character skills,'' \emph{ACM
  Trans. Graph.}, vol.~37, pp. 143:1--143:14, 2018.

\bibitem{Peng2020LearningAR}
X.~Peng, E.~Coumans, T.~Zhang, T.~Lee, J.~Tan, and S.~Levine, ``Learning agile
  robotic locomotion skills by imitating animals,'' \emph{ArXiv}, vol.
  abs/2004.00784, 2020.

\bibitem{Paine2018OneShotHI}
T.~Paine, S.~G. Colmenarejo, Z.~Wang, S.~Reed, Y.~Aytar, T.~Pfaff, M.~W.
  Hoffman, G.~Barth-Maron, S.~Cabi, D.~Budden, and N.~D. Freitas, ``One-shot
  high-fidelity imitation: Training large-scale deep nets with rl,''
  \emph{ArXiv}, vol. abs/1810.05017, 2018.

\bibitem{Xie2018LearningWT}
L.~Xie, S.~Wang, S.~Rosa, A.~Markham, and A.~Trigoni, ``Learning with training
  wheels: Speeding up training with a simple controller for deep reinforcement
  learning,'' \emph{2018 IEEE International Conference on Robotics and
  Automation (ICRA)}, pp. 6276--6283, 2018.

\bibitem{Carlucho2017IncrementalQS}
I.~Carlucho, M.~D. Paula, S.~A. Villar, and G.~G. Acosta, ``Incremental
  q-learning strategy for adaptive pid control of mobile robots,'' \emph{Expert
  Syst. Appl.}, vol.~80, pp. 183--199, 2017.

\bibitem{Pavse2020RIDMRI}
B.~S. Pavse, F.~Torabi, J.~P. Hanna, G.~Warnell, and P.~Stone, ``Ridm:
  Reinforced inverse dynamics modeling for learning from a single observed
  demonstration,'' \emph{IEEE Robotics and Automation Letters}, vol.~5, pp.
  6262--6269, 2020.

\bibitem{Recht2018ATO}
B.~Recht, ``A tour of reinforcement learning: The view from continuous
  control,'' \emph{ArXiv}, vol. abs/1806.09460, 2018.

\bibitem{6386109}
E.~{Todorov}, T.~{Erez}, and Y.~{Tassa}, ``Mujoco: A physics engine for
  model-based control,'' in \emph{2012 IEEE/RSJ International Conference on
  Intelligent Robots and Systems}, 2012, pp. 5026--5033.

\bibitem{Brockman2016OpenAIG}
G.~Brockman, V.~Cheung, L.~Pettersson, J.~Schneider, J.~Schulman, J.~Tang, and
  W.~Zaremba, ``Openai gym,'' \emph{ArXiv}, vol. abs/1606.01540, 2016.

\bibitem{tensorforce}
\BIBentryALTinterwordspacing
A.~Kuhnle, M.~Schaarschmidt, and K.~Fricke, ``Tensorforce: a tensorflow library
  for applied reinforcement learning,'' Web page, 2017. [Online]. Available:
  \url{https://github.com/tensorforce/tensorforce}
\BIBentrySTDinterwordspacing

\bibitem{SpinningUp2018}
J.~Achiam, ``{Spinning Up in Deep Reinforcement Learning},'' 2018.

\bibitem{Kaufmann2020DeepDA}
E.~Kaufmann, A.~Loquercio, R.~Ranftl, M.~M{\"u}ller, V.~Koltun, and
  D.~Scaramuzza, ``Deep drone acrobatics,'' \emph{ArXiv}, vol. abs/2006.05768,
  2020.

\end{thebibliography}


\begin{thebibliography}{1}
\providecommand{\url}[1]{#1}
\csname url@rmstyle\endcsname
\providecommand{\newblock}{\relax}
\providecommand{\bibinfo}[2]{#2}
\providecommand\BIBentrySTDinterwordspacing{\spaceskip=0pt\relax}
\providecommand\BIBentryALTinterwordstretchfactor{4}
\providecommand\BIBentryALTinterwordspacing{\spaceskip=\fontdimen2\font plus
\BIBentryALTinterwordstretchfactor\fontdimen3\font minus
  \fontdimen4\font\relax}
\providecommand\BIBforeignlanguage[2]{{%
\expandafter\ifx\csname l@#1\endcsname\relax
\typeout{** WARNING: IEEEtran.bst: No hyphenation pattern has been}%
\typeout{** loaded for the language `#1'. Using the pattern for}%
\typeout{** the default language instead.}%
\else
\language=\csname l@#1\endcsname
\fi
#2}}

\bibitem{Bertsekas1996NeuroDynamicP}
D.~Bertsekas and J.~Tsitsiklis, ``Neuro-dynamic programming,'' in
  \emph{Encyclopedia of Machine Learning}, 1996.

\bibitem{Levine2018ReinforcementLA}
S.~Levine, ``Reinforcement learning and control as probabilistic inference:
  Tutorial and review,'' \emph{ArXiv}, vol. abs/1805.00909, 2018.

\bibitem{Sutton1998IntroductionTR}
R.~Sutton and A.~Barto, ``Introduction to reinforcement learning,'' 1998.

\bibitem{Kullback1951ONIA}
S.~Kullback and R.~A. Leibler, ``On information and sufficiency,'' \emph{Annals
  of Mathematical Statistics}, vol.~22, pp. 79--86, 1951.

\bibitem{Hornik1991ApproximationCO}
K.~Hornik, ``Approximation capabilities of multilayer feedforward networks,''
  \emph{Neural Networks}, vol.~4, pp. 251--257, 1991.

\end{thebibliography}

\end{document}